\documentclass[journal=jacsat,manuscript=article]{achemso}

\usepackage[version=3]{mhchem} % Formula subscripts using \ce{}

\author{Jia Wang}
\affiliation{Department of Physics, University of Connecticut, Storrs, CT 06269, USA}
\email{wang@phys.uconn.edu}
\author{Chris H. Greene}
\affiliation{Department of Physics, Purdue University, West Lafayette, IN 47907, USA}
\email{chgreene@purdue.edu}

\title[\texttt{achemso} demonstration]{Rydberg states of triatomic hydrogen and deuterium}

\begin{document}
%%%%%%%%%%%%%%%%%%%%%%%%%%%%%%%%%%%%%%%%%%%%%%%%%%%%%%%%%%%%%%%%%%%%%
%% The manuscript does not need to include \maketitle, which is
%% executed automatically.  The document should begin with an
%% abstract, if appropriate.  If one is given and should not be, the
%% contents will be gobbled.
%%%%%%%%%%%%%%%%%%%%%%%%%%%%%%%%%%%%%%%%%%%%%%%%%%%%%%%%%%%%%%%%%%%%%
\begin{abstract}
The triatomic hydrogen ion (H$_3^+$) has spurred tremendous interest in astrophysics in recent decades, and Rydberg states of H$_3$ have also maintained an important role for understanding H$_3^+$ experiments. In a previous study [J. Chem. Phys. \textbf{133}, 234302 (2010)], radiative transitions between neutral H$_3$ Rydberg states were calculated at wavelengths near 7 microns, and could be compared with mid-infrared laser lines observed in hydrogen/rare gas discharges. The present study extends the investigation to wavelengths near 10 -- 13 microns. Rydberg states of D$_3$ are also treated.

\end{abstract}

\section{Introduction}
Although triatomic hydrogen (H$_3$) and its ion (H$_3^+$) are the simplest polyatomic molecules, they have continued to attract intense interest in diverse contexts, ranging from chemistry to astronomy, ever since their discovery. H$_3^+$ plays an important role in astrophysics since it acts as a proton donor in chemical reactions occurring in interstellar clouds \cite{Herbst1973,Watson1973}. Furthermore, this ion also helps to characterize Jupiter's atmosphere from afar \cite{Thompson1989,Owen1993}. H$_3^+$ is the dominant positively charged ion in molecular hydrogen plasmas and was first identified in 1911 by J. J. Thomson with an early form of mass spectrometry \cite{Thomson1911}. Without a stable electronic excited state and a permanent dipole moment, H$_3^+$ cannot be observed by electronic spectroscopy or rotational spectroscopy. Therefore, an infrared rotation-vibration spectrum is the only means to observe this ion. The first observation was carried out by T. Oka in 1980 \cite{Oka1980}. By 2012, more than 600 low-lying rovibrational states of H$_3^+$ had been identified. The good agreement achieved between the experimental spectrum and a first-principles calculation provided a benchmark for calculations on other polyatomic molecules such as water.

One of the biggest surprises among the properties of this simple ion H$_3^+$ is its dissociative recombination (DR) rate, which is important for understanding observations of H$_3^+$ in diffuse interstellar clouds \cite{McCall2003}. Until 2003, the DR process, H$_3^+$+e$^-$$\rightarrow$H$_3$$\rightarrow$ H$_2$+H or H+H+H, has been studied in several different experiments, and had an order of magnitude discrepancy with theoretical expectation at that time. Building on the previous work of Schneider, Orel, and Suzor-Weiner \cite{Orel2000}, Kokoouline and Greene showed \cite{Kokoouline2001, Kokoouline2003} that intermediate Rydberg states of H$_3$ play an important role in the dissociative recombination.  After Rydberg pathways were included in the theoretical description, along with the Jahn-Teller coupling mechanism that excites the vibrational angular momentum mode of the ion, DR theory was able to resolve the discrepancy.  Theory and experiment for this fundamental chemical rearrangement process has now progressed to the point that some energy ranges can even be compared at the level of individual resonance features.\cite{Petrignani2011}  Jungen and Pratt have independently demonstrated\cite{JungenPratt2009} that the overall value of the DR rate coefficient can be accurately determined from a simplified model once the Jahn-Teller capture mechanism is included.

Also in 2003, mid-infrared laser lines at wavelengths near 7 microns in laboratory hydrogen/rare gas supersonic plasmas were observed at Berkeley \cite{Jia2010JCP}. Interestingly, strong IR emission from several massive star-forming regions is observed in a similar wavelength range of the spectrum. Later, these laser lines in the Berkeley experiments were assigned to transitions between metastable H$_3$ Rydberg states, as had been suggested by some detailed theoretical calculations \cite{Jia2010}. A lasing mechanism was also proposed: the population inversion is generated by recombination of the ubiquitous $\rm{H_3^+}$ molecular ion with low-energy electrons. Studies of flowing afterglow plasmas by Glosik et al. suggest a three-body ``collision assisted recombination'' mechanism, rather than a simple two-body process because of the high ($10^{14}$ cm$^{-3}$) He gas density that is present in the supersonic discharge source \cite{Glosik2008}.

More recently, experiments that study lasing in other energy ranges and in systems of other isotopologues such as D$_3$ in similar experimental conditions have been renewed. This has motivated us to extend our previous studies to this wavelength range at around 10--13 micron and to calculate the properties of lasing transitions between the Rydberg states of H$_3$. An extension of our previous study to treat Rydberg states of the other isotopologue D$_3$ is also presented.

\section{Method}
Our theoretical approach to the Rydberg states of H$_3$ is based on multi-channel quantum defect theory (MQDT), one of the most successful techniques for treating Rydberg states in \emph{ab initio} theory. This approach has been detailed in previous work,\cite{Jia2010} so it will only be reviewed briefly here.

In our studies, the model of studying molecular Rydberg energy levels of H$_3$ treats the molecule as a Rydberg electron attached to the H$_3^+$ ion. The interactions between the Rydberg electron and the ion are described by body-frame quantum defects (or the equivalent reaction matrix elements$\tilde K$) that depend on the nuclear geometry. In the MQDT approach, a rovibrational transformation can be applied to construct the lab-frame $K$-matrix using the body-frame quantum defect and the rovibrational wave functions. For $p$-wave Rydberg states, the body-frame quantum defect parameters can be extracted from \emph{ab initio} electronic potential surfaces. For higher orbital angular momentum states ($l>1$), a long-range multipole potential model is adopted. The rovibrational transformation can be formulated as follows:
\begin{equation}
K_{ii'}  = \sum\limits_{\alpha \alpha '} {\left\langle i \right|\left. \alpha  \right\rangle } \tilde K_{\alpha \alpha '} \left\langle {\alpha '} \right|\left. {i'} \right\rangle.
\end{equation}
Here $K_{ii'}$ is an element of the laboratory-frame $K$-matrix, which can be used to solve for eigenenergies $E$ of H$_3$ by solving the following equation, which is the condition to kill exponentially growing components of the wavefunction at $\infty$:
\begin{equation}
\det \left| {\tan \left( {\pi \nu } \right) + K} \right| = 0.
\end{equation}
The laboratory-frame eigenchannels $\left| i \right\rangle$ and the body-frame eigenchannels $\left| \alpha \right\rangle$ are connected by the unitary transformation matrix $U_{i\alpha} = \left\langle i \right|\left. \alpha  \right\rangle$, using the rovibrational wave functions of the H$_3^+$ ion core. To calculate these rovibrational wave functions, an accurate potential energy surface of H$_3^+$ is used,\cite{Cencek1998_1,Cencek1998_2} and the three-body Schr\"{o}dinger equation is solved within the hyperspherical adiabatic representation. In a recent paper,\cite{Jia2010} rovibrational energy levels of H$_3^+$ are calculated and compared with experiment with an accuracy at about 0.2 cm$^{-1}$. Observe that Polyansky and Tennyson achieved an accuracy of 0.02 cm$^{-1}$ using Jacobi coordinates \cite{Tennyson1999}. Their higher accuracy is due to the inclusion of nonadiabatic effects by using different effective reduced masses for vibration and rotation degree of freedom. Because the implementation of their procedure in hyperspherical coordinates is unclear, we have not attempted to reach this higher level of accuracy in the present calculations. However, the permutation symmetry of the rovibrational wave functions can be easily set up in hyperspherical coordinates, which is an important aspect of the rovibrational transformation. Also, the accuracy of the computed Rydberg state energies of H$_3$ is mainly limited by the accuracy of the body frame quantum defects, which yields uncertainties of typically a few cm$^{-1}$. Therefore, the accuracy of the hyperspherical representation is adequate for our present purposes.

The hyperspherical coordinates $\{R,\theta,\varphi\}$ used in our approach are of the Smith-Whitten type,\cite{WhittenSmith1968} which can be defined by the three interparticle distance $r_{12}$, $r_{23}$ and $r_{31}$ through the relations:
\begin{eqnarray}\label{hypercoord1}
r_{12}  &=& 3^{ - 1/4} R\left[ {1 + \sin \theta \sin \left( {\varphi - \pi /6} \right)} \right]^{1/2},\nonumber \\
r_{23}  &=& 3^{ - 1/4} R\left[ {1 + \sin \theta \sin \left( {\varphi  - 5\pi /6} \right)} \right]^{1/2},\\
r_{31}  &=& 3^{ - 1/4} R\left[ {1 + \sin \theta \sin \left( {\varphi  + \pi /2} \right)} \right]^{1/2}. \nonumber
\end{eqnarray}
Together with the Euler angles $\alpha$, $\beta$ and $\gamma$, the three-body system can be described in the body-frame. Similar to the usual Born-Oppenheimer approximation, the adiabatic approach treats the hyperadius $R$ initially as an adiabatic variable, and diagonalizes the Hamiltonian in all other degrees of freedom (such as the hyperangles, Euler angles and spin degrees of freedom) yielding a set of adiabatic potentials and channel functions. The adiabatic corrections and couplings are later included using the ``slow variable discretization'' method \cite{Tolstikhin1996, JiaSVD}.

One of the advantages of adopting this choice for the hyperspherical coordinates is that the basis functions used to discretize the Hamiltonian with the proper permutation symmetry can be easily constructed as,
\begin{equation}
\Phi _{jm_2 K^ +  }^{N^ +  m^ +  \Gamma g_I }  = \frac{{u_j \left( \theta  \right)\left[ {e^{im_2 \varphi } \mathcal{R}_{K^ +  m^ +  }^{N^ +  } \Phi _{g_I }^\Gamma   - \left( { - 1} \right)^{N^ +   + K^ +  } e^{ - im_2 \varphi } \mathcal{R}_{ - K^ +  m^ +  }^{N^ +  } \Phi _{ - g_I }^\Gamma  } \right]}}{{\sqrt {2 + 2\delta _{K^ +  0} \delta _{m_2 0} \delta _{g_I 0} } }},
\end{equation}
where $u_j \left( \theta  \right)$ are a set of fifth-order basis splines which is unaffected by permutations. Here, the rotational part $\mathcal{R}_{K^ +  m^ +  }^{N^ +  }\left({\alpha,\beta,\gamma}\right)$ is given by,
\begin{equation}
\mathcal{R}_{K^ +  m^ +  }^{N^ +  }\left({\alpha,\beta,\gamma}\right)=
\sqrt{\frac{{2N^+ + 1}}{{8\pi ^2 }}} \left[{D^{N^ +}_{m^+ K^+ }\left({\alpha,\beta,\gamma}\right)}\right]^*
\end{equation}
where $D^{N^ +}_{m^+ K^+}$ are the Wigner D functions of the Euler angles. The phase of the Wigner function is chosen as by
Varshalovich \emph{et al}. \cite{Varshalovich} $N^+$ is the total angular momentum of the ion, $K^+$ is the projection of $N^+$ onto the laboratory frame's z-axis, and $m^+$ is the projection onto the body frame's Z-axis. $\Phi^{\Gamma}_{g_I}$ is symmetry-adapted combinations of nuclear-spin functions for three spin half fermions defined as in a previous paper. \cite{Kokoouline2003} $\Gamma = \{A,E\}$ represent the the symmetry representations, where $g_I=0$ for $\Gamma = A$ and $g_I = \pm 1$ (ortho) for $\Gamma = E$ (para). The permutation symmetries for the basis functions chosen for each degree of freedom are shown in \ref{H3spintab}.

\begin{table}
\caption{Permutation symmetry for basis functions of different
degrees of freedom.}\label{H3spintab}
\begin{center}
\begin{tabular}{c|c|c|c}
\hline
%d
Permutation & $e^{im_2\varphi}$ & $\mathcal{R}^{N^ +}_{K^+ m^+}$ & $\Phi^{\Gamma}_{g_I}$\\
Operation& & &\\
\hline
%d
$P_{12}$ & $e^{i4\pi/3}e^{-im_2\varphi}$ & $\left({-}
\right)^{N^+ + K^+}\mathcal{R}^{N^ +}_{-K^+ m^+}$ & $e^{i4\pi g_I/3}\Phi^{\Gamma}_{-g_I}$\\
$P_{23}$ & $e^{i2\pi/3}e^{-im_2\varphi}$ & $\left({-}
\right)^{N^+}\mathcal{R}^{N^ +}_{-K^+ m^+}$ & $e^{i2\pi g_I/3}\Phi^{\Gamma}_{-g_I}$\\
$P_{31}$ & $e^{i2\pi}e^{-im_2\varphi}$ & $\left({-}
\right)^{N^+ + K^+}\mathcal{R}^{N^ +}_{-K^+ m^+}$ & $e^{i2\pi g_I}\Phi^{\Gamma}_{-g_I}$\\
$P_{12}P_{31}$ & $e^{i2\pi/3}e^{im_2\varphi}$ & $\left({-}
\right)^{K^+}\mathcal{R}^{N^ +}_{K^+ m^+}$ & $e^{i2\pi g_I/3}\Phi^{\Gamma}_{g_I}$\\
$P_{12}P_{23}$ & $e^{i4\pi/3}e^{im_2\varphi}$ & $\mathcal{R}^{N^ +}_{K^+ m^+}$ & $e^{i4\pi g_I/3}\Phi^{\Gamma}_{g_I}$\\
\hline
\end{tabular}
\end{center}
\end{table}

Under the condition that $m_2  + g_I  = 3n$ for even $K^+$, and $m_2  + g_I  = 3n+3/2$ for odd $K^+$, it is easy to show that the basis function obeys the permutation symmetry required for three identical fermions:
\begin{subequations}
\begin{equation}\label{permusym1}
P_{12} \Phi _{jm_2 K^ +  }^{N^ +  m^ +  g_I }  =  -
\Phi _{jm_2 K^ +  }^{N^ +  m^ +  g_I },
\end{equation}
and
\begin{equation}\label{permusym2}
\mathcal{A} \Phi _{jm_2 K^ +  }^{N^ +  m^ +  g_I }  =
\Phi _{jm_2 K^ +  }^{N^ +  m^ +  g_I },
\end{equation}
\end{subequations}
where
\begin{equation}\label{permusym3}
\mathcal{A} = 1 - P_{12}  - P_{\,23}  - P_{31}  + P_{12} P_{31}  + P_{12} P_{23}.
\end{equation}

\section{Rydberg transitions of H$_3$ in the 10 -- 13 micron range}
\begin{figure}[htbp]
\includegraphics[width=1.0 \textwidth]{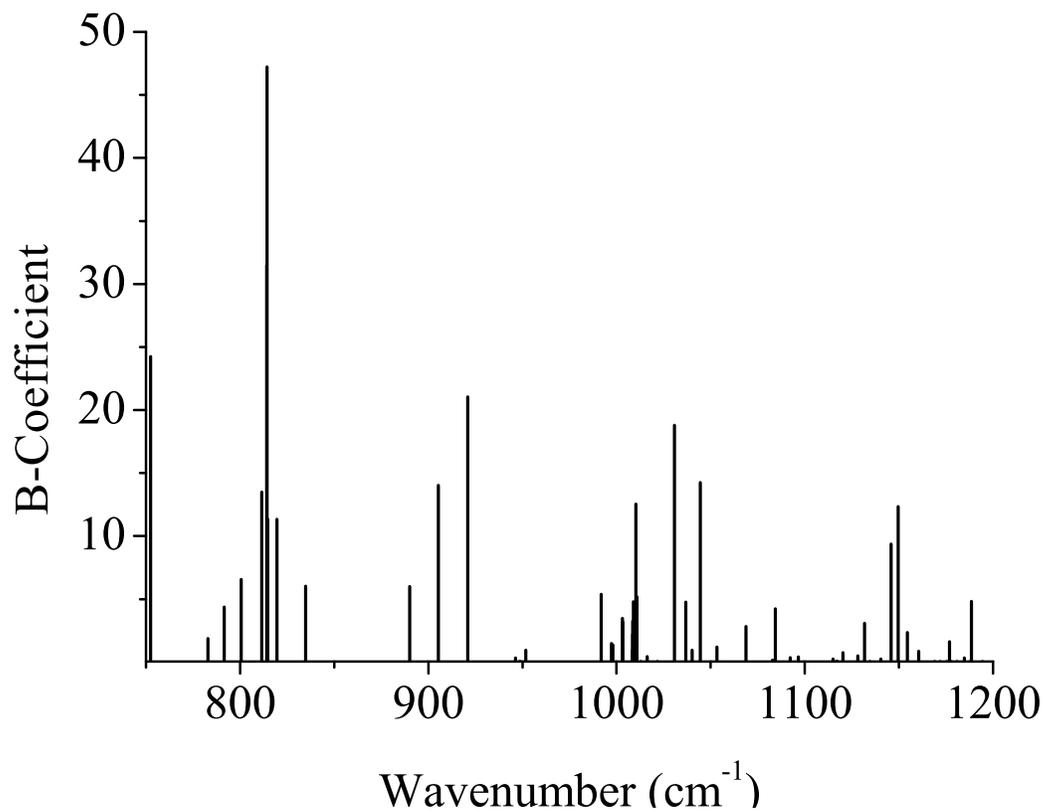}
\caption{Calculated $nd \rightarrow n'p$ transitions of H$_3$ Rydberg states at energy ranging from 750 to 1200 cm$^{-1}$. The y axis shows the theoretical Einstein B-coefficients in units of 10$^{22}$ (m/Js$^2$).}\label{H3lines}
\end{figure}

The method described in last section has been applied to calculate $3p$ and $3d$ Rydberg states of H$_3$, showing good agreement with experiments. The $4d \rightarrow 4p$ and $6d \rightarrow 5p$ Rydberg transitions were used in Ref.\cite{Jia2010JCP} to assign mid-infrared laser lines at wavelengths near 7 microns in laboratory hydrogen/rare gas supersonic plasmas. Here, the Rydberg transitions  near 10--13 microns are calculated and shown in \ref{H3lines}. These transitions are mainly $7d \rightarrow 6p$, $6d \rightarrow 6p$ and $5d \rightarrow 6p$ Rydberg transitions.

\section{$3p_{\pi}$ Rydberg states of D$_3$}
Using the method developed to calculate the Rydberg state energy levels for H$_3$, we have also calculated energy levels for $3p_{\pi}$ Rydberg states of D$_3$. The first step is again calculating the rovibrational states of the ion. In this calculation, the ionic potential surface for D$_3^+$ is adopted from calculations done by Cencek et. al. \cite{Cencek1998_1,Cencek1998_2}, while the same quantum defects as in the case of H$_3^+$ are utilized for the Rydberg state calculation. This should be a good approximation since the quantum defects were calculated under the usual Born-Oppenheimer approximation, where the masses of nucleus are assumed to be infinite. Nevertheless, the rovibrational energy levels of D$_3^+$ are calculated using nuclei mass of deuterium.

A major difference between the calculations of rovibrational states of H$_3^+$ and D$_3^+$ is the different permutational symmetries for the two species: the deuterium nuclei are bosons while the hydrogen nuclei are fermions. The symmetry-adapted combinations of nuclear-spin functions $\Phi^{\Gamma}_{g_I}$ are constructed in the same way as given in by Kokoouline et. al.\cite{Kokoouline2003} Here $\Gamma=\left\{ {A_1, A_2, E}\right\}$ represents the symmetry representations of spin permutation group, and $g_I=0$ for $A_1$ (ortho) and $A_2$ (para) symmetry, while $g_I=\pm1$ for $E$ (meta) symmetry, \cite{Pagani2009} since $E$ representation is two dimensional. The permutation symmetry of these spin functions are tabulated in \ref{D3spintab}.

\begin{table}
  \caption{Permutation symmetry of symmetry-adapted spin functions for D${_3^+}$.}
  \label{D3spintab}
  \begin{tabular}{llll}
    \hline
    Permutation & $\Phi^{A_1}_{g_I}$ (ortho) & $\Phi^{A_2}_{g_I}$ (para)& $\Phi^{E}_{g_I}$ (meta)\\
    Operation & $\left({g_I=0}\right)$  & $\left({g_I=0}\right)$ & $\left({g_I=\pm1}\right)$\\
    \hline
    $P_{12}$ & $e^{i4\pi g_I/3} \Phi^{A_1}_{-g_I}$ & $-e^{i4\pi g_I/3}\Phi^{A_2}_{-g_I}$ & $e^{i4\pi g_I/3}\Phi^{E}_{-g_I}$ \\
    $P_{23}$ & $e^{i2\pi g_I/3} \Phi^{A_1}_{-g_I}$ & $-e^{i2\pi g_I/3}\Phi^{A_2}_{-g_I}$ & $e^{i2\pi g_I/3}\Phi^{E}_{-g_I}$ \\
    $P_{31}$ & $e^{i2\pi g_I} \Phi^{A_1}_{-g_I}$ & $-e^{i2\pi g_I}\Phi^{A_2}_{-g_I}$ & $e^{i2\pi g_I}\Phi^{E}_{-g_I}$ \\
    $P_{12}P_{31}$ & $e^{i2\pi g_I/3} \Phi^{A_1}_{g_I}$ & $e^{i2\pi g_I/3} \Phi^{A_2}_{g_I}$ & $e^{i2\pi g_I/3}\Phi^{E}_{g_I}$ \\
    $P_{12}P_{23}$ & $e^{i4\pi g_I/3} \Phi^{A_1}_{g_I}$ & $e^{i4\pi g_I/3} \Phi^{A_2}_{g_I}$ & $e^{i4\pi g_I/3}\Phi^{E}_{g_I}$ \\
    \hline
  \end{tabular}
\end{table}

The total nuclear-molecular function (including other degree of freedom such as rotation and vibration) should obey the permutation symmetry of three boson system, for example,
\begin{equation}
P_{12} \Phi _{jm_2 K^ +  }^{N^ +  m^ +  \Gamma g_I }  = \Phi _{jm_2 K^ +  }^{N^ +  m^ +  \Gamma g_I },
\end{equation}
and
\begin{equation}
\mathcal{S} \Phi _{jm_2 K^ +  }^{N^ +  m^ +  \Gamma g_I }  = \Phi _{jm_2 K^ +  }^{N^ +  m^ +  \Gamma g_I },
\end{equation}
where $\mathcal{S} = 1 + P_{12}  + P_{23}  + P_{31}  + P_{12} P_{31}  + P_{12} P_{23}$.

Therefore, the basis functions are constructed as,
\begin{equation}
\Phi _{jm_2 K^ +  }^{N^ +  m^ +  \Gamma g_I }  = \frac{{u_j \left( \theta  \right)\left[ {e^{im_2 \varphi } R_{K^ +  m^ +  }^{N^ +  } \Phi _{g_I }^{\Gamma}  + \left( { - 1} \right)^{N^ +   + K^ +  } e^{ - im_2 \varphi } R_{ - K^ +  m^ +  }^{N^ +  } \Phi _{ - g_I }^{\Gamma} } \right]}}{{\sqrt {2 + 2\delta _{K^ +  0} \delta _{m_2 0} \delta _{g_I 0} } }}
\end{equation}
for $\Gamma = A_1$ or $E$, and,
\begin{equation}
\Phi _{jm_2 K^ +  }^{N^ +  m^ +  \Gamma g_I }  = \frac{{u_j \left( \theta  \right)\left[ {e^{im_2 \varphi } R_{K^ +  m^ +  }^{N^ +  } \Phi _{g_I }^{\Gamma}  - \left( { - 1} \right)^{N^ +   + K^ +  } e^{ - im_2 \varphi } R_{ - K^ +  m^ +  }^{N^ +  } \Phi _{ - g_I }^{\Gamma} } \right]}}{{\sqrt {2 + 2\delta _{K^ +  0} \delta _{m_2 0} \delta _{g_I 0} } }}
\end{equation}
for $\Gamma = A_2$, where $g_I$, $m_2$ and $K^+$ satisfies $m_2  + g_I  = 3n$ for even $K^+$, and $m_2  + g_I  = 3n +3/2$ for odd $K^+$.

Using these numerical basis states having the appropriate permutation symmetry, the rovibrational states of D$_3^+$ are calculated and compared with experimental results \cite{Watson1994} in \ref{D3plusErvTab}. The r.m.s. difference between our calculation and experimental results is about $0.11$ cm$^{-1}$.

\begin{table}
  \singlespace
  \caption{Comparison of the calculated infrared transitions with the experimental values \cite{Watson1994}.}\label{D3plusErvTab}
  \scriptsize
  \tabcolsep=0.11cm
  \begin{tabular}{llllllllllll}
  \hline
$\{v_1,v_2,l_2\}'$&$J'$&$G'$&$\Gamma'$&$\{v_1,v_2,l_2\}''$&$J''$&$G''$&$E'$&$E''$&Experiment&This work&Differences\\
& & & & & & & & & (cm$^{-1}$) & (cm$^{-1}$)& (cm$^{-1}$)\\
\hline
$\{0,1,1\}$&1&0&$A_1$&$\{0,0,0\}$&0&0&1887.976&   0.000&1887.976&1888.065&0.089\\
$\{0,1,1\}$&0&1&$E$  & $\{0,0,0\}$&1&1&1834.586&  32.322&1802.263&1802.349&0.085\\
$\{0,1,1\}$&1&0&$A_2$&$\{0,0,0\}$&1&0&1884.308&  43.605&1840.703&1840.789&0.086\\
$\{0,1,1\}$&1&1&$E$  &$\{0,0,0\}$&1&1&1878.488&  32.323&1846.166&1846.256&0.090\\
$\{0,1,1\}$&2&1&$E$  &$\{0,0,0\}$&1&1&1955.905&  32.323&1923.582&1923.674&0.092\\
$\{0,1,1\}$&2&0&$A_2$&$\{0,0,0\}$&1&0&1979.123&  43.605&1935.518&1935.609&0.091\\
$\{0,1,1\}$&2&1&$E$  &$\{0,0,0\}$&1&1&1967.997&  32.323&1935.675&1935.765&0.090\\
$\{0,1,1\}$&1&0&$A_1$&$\{0,0,0\}$&2&0&1887.976& 130.578&1757.397&1757.479&0.082\\
$\{0,1,1\}$&1&1&$E$  &$\{0,0,0\}$&2&1&1878.488&  19.364&1759.124&1759.206&0.082\\
$\{0,1,1\}$&1&2&$E$  &$\{0,0,0\}$&2&2&1847.252&  85.625&1761.627&1761.710&0.083\\
$\{0,1,1\}$&2&1&$E$  &$\{0,0,0\}$&2&1&1955.905& 119.364&1836.540&1836.627&0.087\\
$\{0,1,1\}$&2&0&$A_1$&$\{0,0,0\}$&2&0&1968.201& 130.578&1837.623&1837.712&0.088\\
$\{0,1,1\}$&2&1&$E$  &$\{0,0,0\}$&2&1&1967.997& 119.364&1848.633&1848.720&0.087\\
$\{0,1,1\}$&2&2&$E$  &$\{0,0,0\}$&2&2&1934.967&  85.625&1849.342&1849.430&0.088\\
$\{0,1,1\}$&2&1&$E$  &$\{0,0,0\}$&3&1&1967.997&  49.337&1718.660&1718.739&0.079\\
$\{0,1,1\}$&2&0&$A_2$&$\{0,0,0\}$&3&0&1979.123& 260.450&1718.673&1718.752&0.079\\
$\{0,1,1\}$&2&3&$A_1$&$\{0,0,0\}$&3&3&1880.236& 159.861&1720.375&1720.456&0.081\\
$\{0,1,1\}$&2&3&$A_2$&$\{0,0,0\}$&3&3&1880.262& 159.859&1720.403&1720.481&0.078\\
$\{0,1,1\}$&3&2&$E$  &$\{0,0,0\}$&2&2&2047.093&  85.625&1961.468&1961.567&0.099\\
$\{0,1,1\}$&3&1&$E$  &$\{0,0,0\}$&2&1&2082.031& 119.364&1962.667&1962.762&0.095\\
$\{0,1,1\}$&3&2&$E$  &$\{0,0,0\}$&2&2&2068.516&  85.625&1982.892&1982.988&0.096\\
$\{0,1,1\}$&3&1&$E$  &$\{0,0,0\}$&2&1&2103.337& 119.364&1983.973&1984.068&0.095\\
$\{0,1,1\}$&3&0&$A_1$&$\{0,0,0\}$&2&0&2115.066& 130.578&1984.488&1984.583&0.095\\
$\{0,1,1\}$&3&2&$E$  &$\{0,0,0\}$&3&2&2068.516& 215.904&1852.612&1852.700&0.088\\
$\{0,1,1\}$&3&1&$E$  &$\{0,0,0\}$&3&1&2103.337& 249.337&1854.000&1854.086&0.086\\
$\{0,1,1\}$&3&2&$E$  &$\{0,0,0\}$&3&2&2047.093& 215.904&1831.188&1831.274&0.086\\
$\{0,1,1\}$&3&1&$E$  &$\{0,0,0\}$&3&1&2082.031& 249.337&1832.694&1832.780&0.086\\
$\{0,1,1\}$&3&0&$A_2$&$\{0,0,0\}$&3&0&2093.464& 260.450&1833.014&1833.103&0.089\\
$\{0,1,1\}$&3&3&$A_2$&$\{0,0,0\}$&3&3&2011.611& 159.859&1851.752&1851.841&0.089\\
$\{0,1,1\}$&3&3&$A_1$&$\{0,0,0\}$&3&3&2011.738& 159.861&1851.878&1851.969&0.091\\
$\{0,2,2\}$&0&2&$E$  &$\{0,0,0\}$&1&1&3650.557&  32.323&3618.234&3618.371&0.137\\
$\{0,2,2\}$&1&3&$A_2$&$\{0,0,0\}$&1&0&3646.143&  43.605&3602.538&3602.669&0.131\\
$\{0,2,2\}$&1&2&$E$  &$\{0,0,0\}$&1&1&3694.746&  32.323&3662.423&3662.557&0.133\\
$\{0,2,2\}$&2&4&$E$  &$\{0,0,0\}$&1&1&3662.204&  32.323&3629.881&3630.022&0.141\\
$\{0,2,2\}$&2&3&$A_2$&$\{0,0,0\}$&1&0&3736.252&  43.605&3692.646&3692.785&0.139\\
$\{0,2,2\}$&2&2&$E$  &$\{0,0,0\}$&1&1&3783.078&  32.323&3750.756&3750.879&0.123\\
$\{0,2,2\}$&1&3&$A_1$&$\{0,0,0\}$&2&0&3647.048& 130.578&3516.470&3516.603&0.133\\
$\{0,2,2\}$&1&2&$E$  &$\{0,0,0\}$&2&1&3694.746& 119.364&3575.381&3575.510&0.129\\
$\{0,2,2\}$&1&1&$E$  &$\{0,0,0\}$&2&2&3718.205&  85.625&3632.581&3632.719&0.138\\
$\{0,2,2\}$&2&4&$E$  &$\{0,0,0\}$&2&1&3662.204& 119.364&3542.840&3542.972&0.132\\
$\{0,2,2\}$&2&3&$A_1$&$\{0,0,0\}$&2&0&3733.585& 130.578&3603.006&3603.143&0.137\\
$\{0,2,2\}$&2&2&$E$  &$\{0,0,0\}$&2&1&3783.078& 119.364&3663.714&3663.856&0.142\\
$\{0,2,2\}$&2&1&$E$  &$\{0,0,0\}$&2&2&3806.608&  85.625&3720.984&3721.126&0.142\\
$\{0,2,2\}$&2&4&$E$  &$\{0,0,0\}$&3&1&3662.204& 249.337&3412.867&3412.993&0.126\\
$\{0,2,2\}$&2&3&$A_2$&$\{0,0,0\}$&3&0&3736.252& 260.450&3475.802&3475.923&0.121\\
$\{0,2,2\}$&3&5&$E$  &$\{0,0,0\}$&3&2&3697.465& 215.904&3481.560&3481.688&0.128\\
$\{0,2,2\}$&2&2&$E$  &$\{0,0,0\}$&3&1&3783.078& 249.337&3533.741&3533.873&0.132\\
$\{0,2,2\}$&2&1&$E$  &$\{0,0,0\}$&3&2&3806.608& 215.904&3590.704&3590.836&0.132\\
$\{0,2,2\}$&3&5&$E$  &$\{0,0,0\}$&2&2&3697.464&  85.625&3611.840&3611.984&0.144\\
$\{0,2,2\}$&2&0&$A_2$&$\{0,0,0\}$&3&3&3804.927& 159.859&3645.068&3645.203&0.135\\
$\{0,2,2\}$&2&0&$A_1$&$\{0,0,0\}$&3&3&3805.564& 159.861&3645.704&3645.841&0.137\\
$\{0,2,2\}$&3&4&$E$  &$\{0,0,0\}$&2&1&3794.586& 119.364&3675.222&3675.366&0.144\\
$\{0,2,2\}$&3&3&$A_1$&$\{0,0,0\}$&2&0&3869.806& 130.578&3739.228&3739.376&0.148\\
$\{0,2,2\}$&3&0&$A_1$&$\{0,0,0\}$&3&3&3936.442& 159.861&3776.582&3776.741&0.159\\
$\{0,2,2\}$&3&4&$E$  &$\{0,0,0\}$&3&1&3794.586& 249.337&3545.249&3545.385&0.136\\
$\{0,2,2\}$&3&3&$A_2$&$\{0,0,0\}$&3&0&3864.643& 260.450&3604.194&3604.329&0.135\\
$\{0,2,2\}$&3&2&$E$  &$\{0,0,0\}$&3&1&3915.453& 249.337&3666.116&3666.273&0.157\\
$\{0,2,2\}$&3&1&$E$  &$\{0,0,0\}$&3&2&3939.155& 215.904&3723.251&3723.392&0.141\\
$\{0,2,2\}$&2&0&$A_1$&$\{0,1,1\}$&1&0&3805.564&1887.976&1917.589&1917.646&0.057\\
$\{0,2,2\}$&2&0&$A_1$&$\{0,1,1\}$&2&0&3805.564&1968.201&1837.363&1837.415&0.052\\
$\{0,2,2\}$&2&3&$A_1$&$\{0,1,1\}$&2&3&3733.585&1880.236&1853.349&1853.398&0.049\\
$\{0,2,2\}$&3&1&$E$  &$\{0,1,1\}$&2&1&3911.454&1955.905&1955.549&1955.594&0.045\\
$\{0,2,2\}$&3&0&$A_1$&$\{0,1,1\}$&2&0&3936.442&1968.201&1968.241&1968.304&0.063\\
$\{0,2,2\}$&3&1&$E$  &$\{0,1,1\}$&2&1&3939.155&1967.997&1971.158&1971.238&0.080\\
$\{0,2,2\}$&3&2&$E$  &$\{0,1,1\}$&2&2&3915.453&1934.967&1980.486&1980.581&0.095\\
$\{0,2,2\}$&3&3&$A_1$&$\{0,1,1\}$&3&3&3869.806&2011.738&1858.068&1858.123&0.055\\
$\{0,2,2\}$&3&4&$E$  &$\{0,1,1\}$&3&4&3794.586&1933.565&1861.021&1861.077&0.056\\
$\{0,3,3\}$&1&3&$A_1$&$\{0,2,2\}$&1&3&5513.133&3647.048&1866.084&1866.233&0.149\\
\hline
\end{tabular}
\end{table}

Using these accurate rovibrational states, a rovibrational frame transformation is applied to calculate the 3p$_\pi$ Rydberg states of D$_3$, and compared with experiment results \cite{Watson2003} in \ref{D33ppi}. From this table, the r.m.s. differences between experiment and our calculations are about 6 cm$^{-1}$ for almost all the results here. This might due to the quantum defect surface are optimal for H$_3$, and the accuracy of our result might be improved by simply shifting the quantum defect by a small constant amount.

\begin{table}
  \caption{3p$_{\pi}$ states of D${_3^+}$ comparing with experiment results \cite{Watson2003}.}
  \label{D33ppi}
  \begin{tabular}{llllll}
  \hline
  $N$&$G$&$U$&Experiment&This work&Differences\\
  & & & (cm$^{-1}$) & (cm$^{-1}$) & (cm$^{-1}$)\\
  \hline
    0&1& 1&13040.10&13033.32&-6.78\\
    1&0&-1&13050.01&13048.43&-1.58\\
    1&1& 1&13082.36&13075.96&-6.40\\
    2&0& 1&13146.22&13140.17&-6.05\\
    2&0&-1&13122.60&13121.88&-0.72\\
    2&1&-1&13086.38&13078.91&-7.47\\
    2&1& 1&13167.30&13161.03&-6.27\\
    2&2& 1&13175.00&13168.89&-6.11\\
    2&3& 1&13162.40&13153.10&-9.30\\
    3&-3&1&13288.43&13279.63&-8.80\\
    3&0&-1&13245.79&13239.41&-6.37\\
    3&1&-1&13210.27&13203.07&-7.20\\
    3&1& 1&13295.66&13290.95&-4.71\\
    3&2&-1&13141.43&13130.83&-10.61\\
  \hline
  \end{tabular}
\end{table}
%% classes.  Rather than use \section, an appropriate macro is
%% provided that will always work.
%%%%%%%%%%%%%%%%%%%%%%%%%%%%%%%%%%%%%%%%%%%%%%%%%%%%%%%%%%%%%%%%%%%%%
\acknowledgement
This work has been supported in part by the U.S. Department of Energy, Office of Science. We thank Rich Saykally and his group for discussions relating to this study.

\bibliography{refs}

\end{document}